\documentclass[12pt]{article}
\usepackage{graphicx}
\usepackage{epstopdf}
\usepackage{dcolumn}% Align table columns on decimal point
\usepackage{bm}% bold math
\usepackage{hyperref}
\usepackage{color}

\begin{document}

%%%%%%%%%%%%%%%%%%%%%%%%%%%%%%%%%%%%%%%%%%%

\def\a{\alpha}
\def\b{\beta}
\def\c{\varepsilon}
\def\d{\delta}
\def\e{\epsilon}
\def\f{\phi}
\def\g{\gamma}
\def\h{\theta}
\def\k{\kappa}
\def\l{\lambda}
\def\m{\mu}
\def\n{\nu}
\def\p{\psi}
\def\q{\partial}
\def\r{\rho}
\def\s{\sigma}
\def\t{\tau}
\def\u{\upsilon}
\def\v{\varphi}
\def\w{\omega}
\def\x{\xi}
\def\y{\eta}
\def\z{\zeta}
\def\D{\Delta}
\def\G{\Gamma}
\def\H{\Theta}
\def\L{\Lambda}
\def\F{\Phi}
\def\P{\Psi}
\def\S{\Sigma}

\def\o{\over}
\def\beq{\begin{eqnarray}}
\def\eeq{\end{eqnarray}}
\newcommand{\gsim}{ \mathop{}_{\textstyle \sim}^{\textstyle >} }
\newcommand{\lsim}{ \mathop{}_{\textstyle \sim}^{\textstyle <} }
\newcommand{\vev}[1]{ \left\langle {#1} \right\rangle }
\newcommand{\bra}[1]{ \langle {#1} | }
\newcommand{\ket}[1]{ | {#1} \rangle }
\newcommand{\EV}{ {\rm eV} }
\newcommand{\KEV}{ {\rm keV} }
\newcommand{\MEV}{ {\rm MeV} }
\newcommand{\GEV}{ {\rm GeV} }
\newcommand{\TEV}{ {\rm TeV} }
\newcommand{\1}{\mbox{1}\hspace{-0.25em}\mbox{l}}
\newcommand{\headline}[1]{\noindent{\bf #1}}
\def\diag{\mathop{\rm diag}\nolimits}
\def\Spin{\mathop{\rm Spin}}
\def\SO{\mathop{\rm SO}}
\def\O{\mathop{\rm O}}
\def\SU{\mathop{\rm SU}}
\def\U{\mathop{\rm U}}
\def\Sp{\mathop{\rm Sp}}
\def\SL{\mathop{\rm SL}}
\def\tr{\mathop{\rm tr}}
\def\mpl{M_{\rm Pl}}

\def\IJMP{Int.~J.~Mod.~Phys. }
\def\MPL{Mod.~Phys.~Lett. }
\def\NP{Nucl.~Phys. }
\def\PL{Phys.~Lett. }
\def\PR{Phys.~Rev. }
\def\PRL{Phys.~Rev.~Lett. }
\def\PTP{Prog.~Theor.~Phys. }
\def\ZP{Z.~Phys. }

\def\dd{\mathrm{d}}
\def\ff{\mathrm{f}}
\def\BH{{\rm BH}}
\def\inf{{\rm inf}}
\def\ev{{\rm evap}}
\def\eq{{\rm eq}}
\def\SM{{\rm sm}}
\def\Mpl{M_{\rm Pl}}
\def\GeV{{\rm GeV}}
\newcommand{\Red}[1]{\textcolor{red}{#1}}

%%%%%%%%%%%%%%%%%%%%%%%%%%%%%%%%%%%%%%%%%%%%%%%%%%%%%%%%%%%%%%%
\begin{titlepage}
\begin{center}

\hfill IPMU-14-0212 \\
\hfill ICRR-report-689-2014-15 \\ 
\hfill \today

\vspace{1.5cm}
{{\large\bf 
Phase Locked Inflation } \\
\bf -- Effectively Trans-Planckian Natural Inflation --
}

\vspace{2.0cm}
{\bf Keisuke Harigaya}$^{(a)}$
and
{\bf Masahiro Ibe}$^{(a, b)}$,

\vspace{1.0cm}
{\it
$^{(a)}${Kavli IPMU (WPI), University of Tokyo, Kashiwa, Chiba 277-8583, Japan} \\

$^{(b)}${ICRR, University of Tokyo, Kashiwa, Chiba 277-8582, Japan}
}

\vspace{2.0cm}
\abstract{
A model of natural inflation with an effectively trans-Planckian decay constant can be 
easily achieved by the ``phase locking" mechanism while keeping field values in 
the effective field theory within the Planck scale.
We give detailed description of  ``phase locked" inflation based on this mechanism. 
We also construct supersymmetric natural inflation based on this mechanism and show 
that the model is consistent with low scale supersymmetry.
We also investigate couplings of the inflaton with the minimal supersymmetric standard model
to achieve an appropriate reheating process.
Interestingly, in a certain class of models, we find that 
the inflation scale is related to the mass of the right-handed neutrino 
in a consistent way with the seesaw mechanism.
}
\end{center}
\end{titlepage}
\setcounter{footnote}{0}

%%%%%%%%%%%%%%%%%%%
%---------------SECTION-------------------%
%%%%%%%%%%%%%%%%%%%
\section{Introduction}
Cosmic inflation~\cite{Guth:1980zm} (see also Ref.~\cite{Kazanas:1980tx}) is a very successful paradigm
of modern cosmology which not only solves various problems of the standard 
cosmology such as the horizon problem and the flatness problem, but 
also provides the origins of the large scale structure of the universe 
and the fluctuation of the cosmic microwave background (CMB) radiation~\cite{Mukhanov:1981xt}.
Precise measurements of the CMB~\cite{Hinshaw:2012aka,Ade:2013uln}
have shown that the so-called slow-roll inflation~\cite{Linde:1981mu,Albrecht:1982wi}
is remarkably successful. 

Among various models of cosmic inflation, so-called natural inflation\,\cite{Freese:1990rb} 
has been considered to be one of the most attractive models, in which a pseudo-Nambu-Goldstone (NG) boson
(pseudo-NGB)
associated with spontaneously breaking of an approximate symmetry plays a role of the inflaton.
The shape of the inflaton potential is well controlled by small explicit breaking of the spontaneously
broken symmetry.
After the announcement of a large tensor fraction in the CMB, $r=O(0.1)$, by the BICEP2
collaboration~\cite{Ade:2014xna},
natural inflation has received renewed attention since it
shows a perfect fit with the BICEP2 result\,\cite{Freese:2014nla}.

Worrisome features of natural inflation are, however, the requirements of a decay constant
larger than the Planck scale and the trans-Planckian variation of the inflaton field during inflation.
Such trans-Planckian decay constant and  field variation are not easy to be justified 
within the framework of the effective field theory valid below the Planck scale. 
There the Lagrangian is at the best given by a series expansion of fields 
with higher dimensional operators suppressed by the Planck scale.
In this view point of the effective field theory, the trans-Planckian decay constant and the field variation seem 
inevitably sensitive to the theory beyond the Planck scale.

In order to avoid troubles of Planck suppressed operators, so far, many attempts 
have been made where the trans-Planckian decay constant and field variation are realized
in effective ways while keeping actual field values within the Planck scale.
For example, alignment between potentials of natural inflation generated by multiple explicit breaking
effects leads to the one with an effective decay constant  larger than 
the Planck scale~\cite{Kim:2004rp} (see also Ref.\,\cite{Choi:2014rja,Higaki:2014pja} for recent discussions).
In $N$-flation~\cite{Dimopoulos:2005ac}, many pseudo-NGBs are introduced and their collective dynamics lead to an effectively large decay constant.
The monodromic behaviors of the pseudo-NGB also
% make it possible to
realize models with an effective decay constant larger than the Planck scale, 
which is originally proposed in models based on string theory~\cite{Silverstein:2008sg,McAllister:2008hb}
and are generalized in field theoretic approaches\,\cite{Kaloper:2008fb,Berg:2009tg,Kaloper:2011jz,Dine:2014hwa}.
%(See also Refs.\,\cite{???,Harigaya:2014eta} for recent developments.)
It is also pointed out that natural inflation potential generated by an anomaly in a large $N_c$ gauge dynamics
% is also pointed out to
has an effective decay constant much larger than the scale of the spontaneous symmetry breaking\,\cite{Yonekura:2014oja}.

In Ref.\,\cite{Harigaya:2014eta}, the authors  proposed another class of models which 
achieve trans-Planckian field variations in an effective way in a field theoretic approach.
There, the inflaton potential exhibits monodromic behavior due to the ``phase locking" mechanism 
between two pseudo-NGBs, which can be immediately applied to natural inflation.
In this paper, we refer to models in this class as  ``phase locked" inflation,
and give detail discussion on the models.

We also propose a supersymmetric model of natural inflation based on 
the ``phase locking" mechanism.
In particular, we show that it is possible to construct a model consistent with  
low-scale supersymmetry  in our approach. 
It should be emphasized that in the most models of supersymmetric natural inflation so far proposed,
the explicit symmetry breaking term 
leading to the inflaton potential is proportional to the constant term in the superpotential, 
i.e. the gravitino mass. 
Therefore, in order to reproduce the measured properties of the CMB,  
the gravitino mass turns out to be very large and are not compatible 
with low scale supersymmetry in a multi-TeV range.
In our model, on the other hand, the explicit symmetry breaking is separated from 
the $R$-symmetry breaking, and hence, the model does not 
require a large gravitino mass even for a high scale inflation.
 
The organization of  this paper is as follows.
In the next section, we give detailed descriptions of generic features of 
natural inflation based on the ``phase locking" mechanism.
We also discuss details of the inflation dynamics such as the initial condition of the inflaton.
In section\,\ref{sec:SUSY}, we show a simple model of the supersymmetric natural inflation 
based on ``phase locked" inflation which is compatible with low scale supersymmetry breaking. 
We also discuss the reheating process and some interesting features of the model. 
The final section is devoted to discussion and conclusions.

%%%%%%%%%%%%%%%%%%%
%---------------SECTION-------------------%
%%%%%%%%%%%%%%%%%%%
\section{Phase Locked Inflation}
\label{sec:locking}
In this section, we give a detailed description of 
the ``phase locking" mechanism proposed in Ref.\,\cite{Harigaya:2014eta}, which 
can be applied to natural inflation so that the model has an effective decay
constant larger than the Planck scale.

\subsection{Brief review on natural inflation}
Before going to details of the mechanism, let us briefly review natural inflation~\cite{Freese:1990rb}.
In natural inflation, the inflaton is identified with a pseudo-NGB
associated with  spontaneously breaking of an approximate symmetry.
When the explicit breaking of the symmetry is dominated by a single breaking term,
the inflaton potential is given by,
\begin{eqnarray}
\label{eq:NI}
 V = \Lambda^4 (1 - \cos(a/f)) \ ,
\end{eqnarray}
where $\L$ denotes a parameter encoding the explicit 
breaking while $f$ is the decay constant of the pseudo-NGB $a$.
From this potential, we immediately find that the slow-roll parameters, 
\begin{eqnarray}
\epsilon &=& \frac{\mpl^2}{2}\left(\frac{V'}{V}\right)^2 = \frac{\mpl^2}{2f^2}\frac{\sin^2(a/f)}{(1-\cos(a/f))^2}\ , \cr
\eta &=& \frac{\mpl^2V''}{V} =  \frac{\mpl^2}{f^2}\frac{\cos(a/f)}{(1-\cos(a/f))}\ ,
\end{eqnarray}
satisfy the slow-roll conditions, $\epsilon \ll 1$  and $|\eta| \ll 1$, only for
\begin{eqnarray}
f  \gg \mpl \quad  {\rm and}\quad |a|  \gg \mpl  \ .
\end{eqnarray}
Thus, if we simply identify the decay constant $f$ with the scale of spontaneous symmetry breaking by a vacuum expectation 
value (VEV) , i.e.
\begin{eqnarray}
 X  \sim O(f) e^{i a/f}\ ,
\end{eqnarray}
where $X$ denotes a field which is responsible for spontaneous symmetry breaking, 
the large decay constant requires a VEV much larger than the Planck scale.
As cautioned in introduction, such a VEV larger than the Planck scale is difficult to be 
justified in the effective field theory below the Planck scale.
We loose control over the inflaton potential from contributions of higher dimensional
operators suppressed by the Planck scale.

\subsection{Effectively large decay constant by phase locking}
The decay constant $f$ is, however, not necessary to be of the order of the scale of spontaneous symmetry
breaking.
In fact, a decay constant much larger than the VEVs can be realized in a very simple manner,
the ``phase locking" mechanism  proposed in Ref.\,\cite{Harigaya:2014eta}. 

To illustrate the mechanism, let us consider a model with an approximate $U(1)$ symmetry under which 
two complex scalar fields $\phi$ and $S$ have charges of $N$ and $1$, respectively. 
The scalar potential consistent with the $U(1)$ symmetry is given by
\begin{eqnarray}
V = V(\phi \phi^*, S S^*, \phi^* S^N) \ .
\end{eqnarray}

Now, let us assume that both $\phi$ and $S$ obtain VEVs in a similar size.
In such a case, we have two candidates of the NGB, i.e. the phases of $\phi$ and $S$; 
\begin{eqnarray}
\phi|_{\rm NGB} = |\vev\phi|\, e^{i \arg\phi}\ , \quad
S|_{\rm NGB} = |\vev S|\, e^{i \arg S}\ .
\end{eqnarray}
Then, due to terms depending on $\phi^* S^N$, one of the 
NGB candidates obtains a large mass from%
\footnote{Here, we suppressed phases of the potential terms for simplicity.}
\begin{eqnarray}
\label{eq:lock}
 V \propto \phi^* S^N + h.c. = 2 |\vev\phi | |\vev{S}|^N \cos( \arg \phi - N\arg S)\ .
\end{eqnarray}
As a result, the remaining NGB which will be identified with the inflaton 
corresponds to a combination,
\begin{eqnarray}
  a \propto N\arg\phi + \arg S \ .
\end{eqnarray}
Putting all together, the canonically normalized NGB, $a$, is given by,
\begin{eqnarray}
\label{eq:NGB}
\phi|_{\rm NGB} &=& |\vev\phi|\, e^{i N a/f}\ , \nonumber\\
S|_{\rm NGB} &=& |\vev S|\, e^{i a/f}\ \ , \nonumber\\
f &\equiv& \sqrt{2 N^2 |\vev{\phi}|^2 + 2 |\vev{S}|^2}\ .
\end{eqnarray}
Remarkably, the decay constant $f$ can be much 
larger than the VEVs $\vev{\phi} \simeq \vev{S}$ for $N\gg 1$.

To generate the inflaton potential as in Eq.\,(\ref{eq:NI}), 
we softly break the $U(1)$ symmetry by 
\begin{eqnarray}
\label{eq:breaking}
V = M^3 S + {\rm h.c.}\ .
\end{eqnarray}
Here, one may consider that the explicit breaking parameter $M^3$ is a spurious field whose $U(1)$ charge is $-1$.
The inflaton potential is dominated by the contribution of this explicit breaking term.
Substituting Eqs.~(\ref{eq:NGB}) to (\ref{eq:breaking}), we obtain the desirable  inflaton potential,%
\footnote{
For a very large $N$, the scalar potentials which lock the 
phases of $\phi$ and $S$, i.e. the $\phi^* S^N$-dependent terms, 
are highly suppressed due to their high mass dimensions. 
In this case the potential in Eq.\,(\ref{eq:lock}) is highly suppressed and hence the mode
proportional to $\arg \phi - N \arg S$ becomes lighter than $a$ which ruins our discussion\,\cite{Harigaya:2014eta}.
A possible solution to this problem is discussed at the end of this section.
}
\begin{eqnarray}
\label{eq:lockedNI}
V(a) = 2 |M^3 \vev{S}| \left( 1- {\cos}(a/f)\right)\ .
\end{eqnarray}
Here, we have eliminated a constant phase and a sign inside the cosine by shifting $a$,
and added a constant term to the potential so that the cosmological constant vanishes at the vacuum.
In this way, we obtain a model of natural inflation with a decay constant much larger than
the scale of symmetry breaking.

\begin{figure}[tb]
\begin{center}
  \includegraphics[width=1\linewidth]{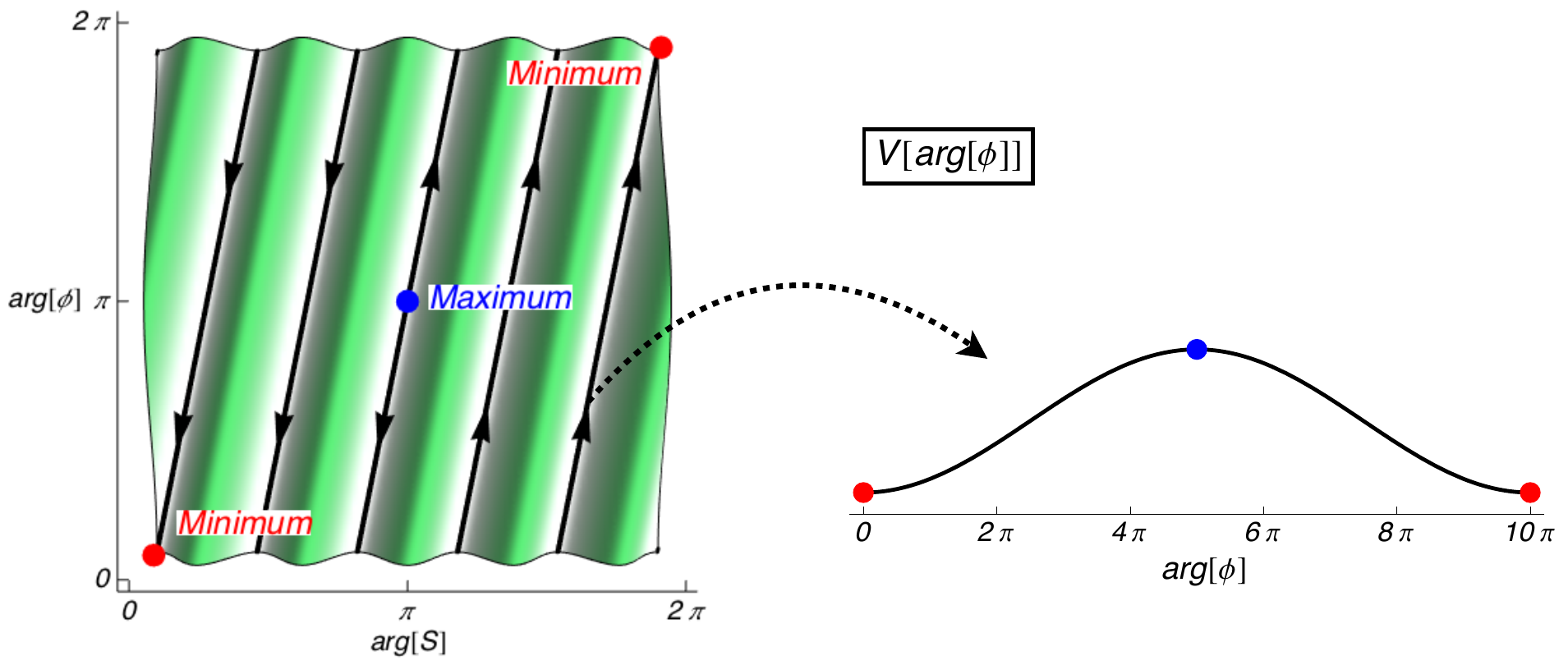}
 \end{center}
\caption{\sl \small
(Left) A scalar potential as a function of the phases of $S$ and $\phi$ for $N=5$.
The inflaton trajectory corresponds to the valley is shown by the solid line.
Arrowheads indicate the height of the inflaton potential along the trajectory,
and the field values which maximize and minimize the potential 
in Eq.\,(\ref{eq:lockedNI}) are denoted by points.
(Right) The shallow inflaton potential in Eq.\,(\ref{eq:lockedNI}) along the valley in the left panel 
as a function of $\arg[\phi]$.
}
\label{fig:lock}
\end{figure}

It should be noted that the potential term Eq.\,(\ref{eq:lock}) tightly interrelates 
the phases of $\phi$ and $S$ so that
\begin{eqnarray}
\arg S = \frac{1}{N}\arg \phi\ , \quad {\rm i.e.}\quad  S \propto \phi^{1/N}\ ,
\end{eqnarray}
in the effective theory below the breaking scale.
We refer to this phenomenon as ``phase locking".
In view of the phase locking mechanism, the effectively large decay constant 
can be understood in the following way.
Due to the phase locking, when $\phi$ rotates $2\pi$, $S$ rotates only $2\pi/N$.
Since the inflaton potential is provided by the explicit breaking term on the phase rotation symmetry
of $S$  as in Eq.\,(\ref{eq:breaking}),
the periodicity of the potential in terms of the phase of $\phi$ is effectively enlarged to $2\pi N$.
Besides, for  a large $N$, the pseudo-NGB $a$ is mainly composed of the phase of $\phi$ due 
to its large charge $N$ as long as $\vev \phi \sim \vev S$.
Therefore, the inflaton dynamics is approximately identified with the dynamics of the phase of $\phi$, 
which can take effectively larger field values than the VEVs of $\phi$ and $S$.
In Fig.~\ref{fig:lock}, we show the inflaton trajectory on the ($\arg S$, $\arg\phi$) plane for $N=5$.%
\footnote{
For large $N$, distances between the valleys of the potential is small and hence the tunneling process between valleys might disturb the inflation dynamics (see Fig.~\ref{fig:tunneling} in Appendix).
We estimate the tunneling rate in Appendix and find that the tunneling process is irrelevant for sufficiently large 
$\vev{\phi}$ and $\vev{S}$.
}

The explicit breaking of the $U(1)$ symmetry may also be provided by non-perturbative dynamics.
For example, let us assume a QCD-like theory which exhibits spontaneous breaking 
of chiral symmetries at a dynamical scale $\Lambda_{\rm dyn}$ far below the VEVs of $\phi$ and $S$.
We couples $S$ to the gauge theory via a yukawa coupling,
\begin{eqnarray}
{\cal L}_{\rm int} = y S Q\bar{Q},
\end{eqnarray}
where $Q$ and $\bar{Q}$ are fermion fields charged under the gauge symmetry and $y$ is a constant.
Since the $U(1)$ symmetry is now anomalous, the scalar potential of the pseudo-NGB $a$ is generated.

We first consider the case where there are fermions which are charge under the gauge symmetry and are lighter than $\Lambda_{\rm dyn}$.
In this case, as is the case with the QCD-axion~\cite{Peccei:1977hh,Weinberg:1977ma,Wilczek:1977pj}, the potential of $a$ is given by
\begin{eqnarray}
\label{eq:pot fermion}
V &\simeq& m_f\Lambda_{\rm dyn}^3 \left(1- \cos\left(\arg S\right)\right)\ , \nonumber\\
&\simeq& m_f\Lambda_{\rm dyn}^3\left(1- \cos( a/f)\right)\ .
\end{eqnarray}
where $m_f$ is the mass of the light fermions.
Here, we have assumed that the light fermions have masses of the same order, for simplicity.

Next, let us consider the case where there is no light fermion which is charged under the gauge symmetry.
In this case, it is difficult to calculate the scalar potential of $a$; the simple cosine form such as the one in Eq.~(\ref{eq:pot fermion}) is not guaranteed.
For large $N_c$ dynamics, however, the potential is expanded as~\cite{Yonekura:2014oja,Witten:1980sp}
\begin{eqnarray}
V (a) = \Lambda_{\rm dyn}^4\frac{a^2}{f^2} \left( 1 + {\cal O} \left(\frac{a^2}{f^2 N_c^2}\right) \right).
\end{eqnarray}
For $a<f N_c$, the potential is approximated by a quadratic one.
We can say that the effective decay constant is further enhanced by $N_c$,
\begin{eqnarray}
f_{\rm eff} \simeq N_c f \simeq N_c N  |\vev{\phi}|.
\end{eqnarray}
This enhancement is also helpful to have a successful natural inflation model from small field values~\cite{Yonekura:2014oja}.

Before closing this section, let us discuss an issue for a very large $N$.
In this case, the phase locking potential in Eq.\,(\ref{eq:lock}) is highly suppressed due to their high mass dimensions, 
and hence, the other inflaton candidate, $b\propto \arg \phi - N \arg S$, becomes lighter 
than $a$ for a given $M^3$ 
unless the VEVs of $S$ and $\phi$ are very close to $M_{\rm PL}$.
Unfortunately, however, $b$ cannot be a good candidate of the inflaton since
the effective decay constant of $b$ is given by $f/N$, and hence, 
we cannot achieve an effectively large decay constant for $b$.

This problem can be evaded if we generalize the idea to a generic approximately $U(1)$ symmetric model
and introduce complex scalar fields $\phi_{i}$ with $U(1)$ charges of ${q_i}$~$(i = 1,2,\cdots,~|q_i|\leq |q_{i+1}|,~q_1 = 1)$.
We further assume that the $U(1)$ symmetry is spontaneously broken by the VEVs of all the 
complex scalar fields in similar sizes.
With this generalization, the potential terms which corresponds to the phase locking potential in Eq.\,(\ref{eq:lock})
are not necessarily suppressed since we have fields with variety of charges.
Therefore, in this case, we expect that the pseudo-NGB $a$ associated with the spontaneous symmetry breaking 
resides in the complex scalar fields; 
\begin{eqnarray}
\label{eq:NGBgen}
\phi_{i} &\rightarrow& |\vev{\phi_{i}}| {\rm exp}\left[i q_i \frac{a}{f}\right],\nonumber\\
f &\equiv& \sqrt{2 \sum_i q_i^2 |\vev{\phi_i}|^2}\ ,
\end{eqnarray}
while other phase directions are strongly fixed by the phase locking potentials.
The larger charge the field has, the more pseudo-NGB component the field has.
Then, by explicitly breaking the $U(1)$ symmetry by a linear term of the field 
with the smallest charge, the inflaton potential is that of natural inflation with the decay constant $f$, 
which can be much larger than a VEV of each field.

\subsection{Inflaton Dynamics and Initial Condition}
Finally, let us discuss how inflation begins in our set up.
Since we assume that  the VEVs of $\phi$ and $S$ are below the Planck scale,
it is natural to expect that the universe is in a symmetric phase at a very early epoch at around 
the Planck time. 
Then, it is generically expected that cosmic strings are formed when the universe experiences 
the $U(1)$ phase transition.%
\footnote{Here, the cosmic strings corresponds to the so-called global string which
is expected to be stable if it stretches across the Hubble volume.
 }
Then, as the energy of the universe further drops, the explicit breaking of the $U(1)$ symmetry
becomes important, and domain walls are formed in between cosmic strings.
As a result, the universe is dominated by networks of strings and domain walls.%
\footnote{We assume that the universe is open and enough long-lived; otherwise the universe collapses 
before entering this phase.}
This situation is quite similar to the evolution of the string-domain wall network in the case of 
the QCD-axion~\cite{Sikivie:1982qv} which is confirmed by numerical simulations\,\cite{Ryden:1989vj}.

In the string-domain wall network, we can find domain walls surrounded by cosmic strings 
around which the phases of $\phi$ and $S$ on the inflaton trajectory shown in Fig.~\ref{fig:lock} wrap.
The energy density of this type of the domain walls is as large as $\Lambda^4$.%
\footnote{
There also exist domain walls surround by cosmic strings around which 
the phases of $\phi$ and $S$ not on the inflaton trajectory wrap.
The energy density of this type is as large as $|\vev{\phi}||\vev{S}^N|/M_{\rm Pl}^{N-3}$.
}
A typical thickness of domain walls is as large as $m_a^{-1}$, where $m_a$ is the mass of the inflaton~\cite{Linde:1994hy,Vilenkin:1994pv}.
On the other hand, the Hubble radius of a domain wall-dominated universe is as large as $\mpl/ (m_a f)$.
For $f> \mpl$, domains walls are thicker than the Hubble radius of the domain wall dominated universe.
Thus, a region inside a domain wall can be regarded as an independent universe due to causality. 
Well inside domain walls, the field value of the inflaton is close to the ``Maximum" in Fig.\,\ref{fig:lock},
where the slow-roll conditions are satisfied.
That is, natural inflation begins in a region well inside the domain wall.
This situation is the same as topological inflation~\cite{Linde:1994hy,Vilenkin:1994pv}.

%Before closing this section, let us comment on the relation between the phase locking mechanism and the 

%%%%%%%%%%%%%%%%%%%
%---------------SECTION-------------------%
%%%%%%%%%%%%%%%%%%%
\section{Supersymmetric Natural Inflation}
\label{sec:SUSY}

In this section, we propose a simple model of supersymmetric natural inflation compatible with low scale supersymmetry breaking, where the effectively trans-Planckian decay constant is realized by the phase locking mechanism.

\subsection{Phase locking and NG multiplet}
According to the  recipe to construct  models with the phase locking mechanism, 
we assume that the model possesses an approximate $U(1)$ symmetry where the explicit 
breaking leads to the inflaton potential.
Concretely, we introduce four chiral multiplets $\phi$, $\bar{\phi}$, $S$ and $\bar{S}$ with $U(1)$ charges of $+N$, $-N$, $+1$ and $-1$, respectively.
We also assume that the model possesses the $R$-symmetry under which the above chiral multiplets are neutral.

In order for them to obtain non-zero VEVs and to have their phases locked with each other,
we further introduce three chiral multiplets, $Y_1$, $Y_2$ and $Y_3$ with the $R$ charges of $2$ 
but with vanishing $U(1)$ charges.
With these charge assignments, the generic superpotential is given by
\begin{eqnarray}
W = \sum_{i=1,2,3} Y_i F_i (\phi \bar{\phi},S\bar{S}, S^N\bar{\phi},\bar{S}^N \phi),
\end{eqnarray}
where $F_i$ are generic holomorphic functions.
With this superpotential, all directions except for the pseudo-NG multiplet are fixed, and the phases are locked.

%%%%%%%%%%%%%%%%%%%%%%%%%%%%%%%%%%%%%%%%%%%%%%%%
\begin{table}[tdp]
\begin{center}
\begin{tabular}{|c|ccccccccc|}
\hline
& $\phi$ & $\bar{\phi}$ & $S$ & ${\bar{S}}$ & $Y_{1,2}$ & $Y_3$ & $X$ & $\epsilon$ & $\bar{\epsilon}$ \\
\hline
$U(1)$ & $+N$ & $-N$ & $+1$ & $-1$ & $0$ & $0$ & $0$ & $+1$ & $-1$ \\
$U(1)_R$ & $0$ & $0$ & $0$ & $0$ & $2$ & $2$ & $2$ & $0$ & $0$\\
$Z_2$ & $(-)^{N+1}$ & $(-)^{N+1}$ & $-$ & $-$ & $+$ & $-$ & $-$ & $+$ & $+$ \\
$Z_{2C}$ & 
$\phi\leftrightarrow \bar{\phi}$ 
&  
$\bar\phi\leftrightarrow {\phi}$ 
& $S\leftrightarrow \bar{S}$ 
& $\bar S\leftrightarrow S$   &&  & & $\epsilon \leftrightarrow \bar{\epsilon}$ 
&$\bar\epsilon \leftrightarrow {\epsilon}$  
\\
\hline
\end{tabular}
\end{center}
\caption{\sl \small Charge assignment of the inflaton sector.}
\label{tab:charge1}
\end{table}%
%%%%%%%%%%%%%%%%%%%%%%%%%%%%%%%%%%%%%%%%%%%%%%%%

For simplicity, we further introduce discrete symmetries $Z_2\times Z_{2C}$ symmetry whose charge assignments are 
given in Table\,\ref{tab:charge1}.
The generic leading order superpotential is given by
\begin{eqnarray}
\label{eq:superpotential}
W = Y_1 ( \lambda_{1\phi}\phi \bar{\phi} - v_1^2 ) + Y_2 ( \lambda_{2\phi} \phi \bar{\phi} + \lambda_{2S} S \bar{S}- v_2^2 ) + \lambda_3 Y_3(S^N \bar{\phi} + \bar{S}^N \phi)/\mpl^{N-1},
\end{eqnarray}
where $\lambda_{1\phi}$, $\lambda_{2\phi}$, $\lambda_{2S}$ and $\lambda_{3}$ denote dimensionless coupling constants while $v_{1,2}$ are dimensional constants
in a similar size with each other.
We have eliminated the coupling of $Y_1$ with $S \bar{S}$ by taking appropriate linear combinations of $Y_1$ and $Y_2$.
We take $\lambda_{1\phi}$, $v_1^2$, $\lambda_{2S}$ and $v_2^2$ to be real and positive without loss of generality.
As we will discuss later, we find the $Z_2 \times Z_{2C}$ symmetry is helpful to suppress the decay of the inflaton into gravitinos, though not mandatory.
It should be emphasized that the other higher dimensional operators suppressed by $\mpl$ are not very relevant 
for the following discussion since we assume that the VEVs of $\phi$'s and $S$'s
%, i.e. $v_{\phi,S}$
are below the Planck scale.

Now, let us extract a multiplet of the pseudo-NGB  associated with the $U(1)$ symmetry breaking.
First, with the $F$ term conditions of $Y_1$ and $Y_2$, the chiral multiplets $\phi$, $\bar{\phi}$, $S$ and $\bar{S}$ are reduced to
\begin{eqnarray}
\phi \rightarrow v_\phi e^{\rho}\ ,\quad\bar{\phi} \rightarrow v_{\phi} e^{-\rho}\ ,\nonumber\\
S \rightarrow v_S e^{\sigma}\ , \quad \bar{S} \rightarrow v_S e^{-\sigma}\ ,
\end{eqnarray}
where $v_{\phi} = v_1 / \sqrt{\lambda_{1\phi}}$ and $v_S = v_2/\sqrt{\lambda_{2S}}$, and $\rho$ and $\sigma$ are
the chiral multiplets.
%Here, we have assumed that $v_\phi$ and $v_S$ is smaller than the Planck scale and hence supergravity effect is negligible.
Then, the $F$ term condition of $Y_3$ further requires
\begin{eqnarray}
\label{eq:lockingSUSY}
e^{N\sigma -\rho } + e^{-N\sigma + \rho} =0 \  \Longleftrightarrow \ \sigma = \frac{\rho}{N} + i \pi \frac{2n +1}{2N}\ ,(n=0,1,\cdots,2N-1)\ ,
\end{eqnarray}
which is nothing but the phase locking.
In the following, we fix the phase locking direction to $\sigma = \rho/N + i \pi / (2N)$
by rotating $S$ and $\bar{S}$ by $e^{- i\pi n/N}$ and $e^{+ i\pi n/N}$.%
\footnote{
Note that the superpotential in Eq.~(\ref{eq:superpotential}) is invariant under this rotation by further rotating $Y_3$ by $-1$.}

According to the discussion in the previous section, we  identify $\rho$ as the main component of 
the pseudo-NG multiplet for a large $N$ as long as $v_{\phi} \sim v_{S}$.
The $F$-term condition of $Y_3$ generates the non-trivial potential mainly to $\sigma$.%
\footnote{We may explicitly extract the massless chiral multiplet in the diagonalized base of $\rho$ and $\sigma$,
although the following discussion is not significantly altered.}
With this identification, we refer to $\rho$ as the inflaton multiplet
whose imaginary part of the scalar component plays a role of the inflaton in natural inflation.
We also refer to the real part of the inflaton multiplet as the s-inflaton.

\subsection{Inflation dynamics}
In order to give a potential to the inflaton via  explicit $U(1)$ breaking,
we introduce a chiral multiplet $X$ and spurious fields $\epsilon$ and $\bar{\epsilon}$ whose charge assignments are given in Table~\ref{tab:charge1}.
We assume that the $Z_{2C}$ symmetry is not explicitly broken by these fields, and hence,
we fix $ \epsilon = \bar{\epsilon}$.
By using these multiplet, we introduce explicit $U(1)$ breaking terms,
\begin{eqnarray}
\label{eq:breakingSUSY}
\Delta W &=& X (\bar{\epsilon} S + \epsilon \bar{S} ) = \epsilon X (S + \bar{S}) \ .
%\nonumber\\
% &=& \epsilon v_S X \left(
% {\rm exp}\left[ \frac{b}{N} + i \pi \frac{2n+1}{2N} \right] +  {\rm exp}\left[-\frac{b}{N} - i \pi \frac{2n+1}{2N} \right] 
% \right)
\end{eqnarray}
By substituting the NG modes appeared in Eqs.\,(\ref{eq:lockingSUSY}) and (\ref{eq:breakingSUSY}), 
we obtain the superpotential of the pseudo-NG multiplet,
\begin{eqnarray}
\label{eq:supereff}
\Delta W 
&=& \epsilon v_S X \left(
{\rm exp}\left[ \frac{\rho}{N} + i \frac{\pi}{2N} \right] +  {\rm exp}\left[-\frac{\rho}{N} - i \frac{\pi}{2N} \right] 
 \right),
\end{eqnarray}
which has a similar structure with the superpotential used in the supergravity chaotic inflation~\cite{Kawasaki:2000yn}.%
\footnote{
The resultant superpotential 
is analogous to the one given in Ref.~\cite{Kallosh:2014vja} to construct natural inflation, which in our terminology
is given by  $W \propto X \sin[\rho/N]$. 
It should be emphasized that our model not only provides a simple ultra-violet completion to their model 
but also realizes the large decay constant via the phase locking mechanism at the same time.
}
From this superpotential, we again find that the defining region of the imaginary part of $\rho$, i.e. the inflaton,
is effectively enhanced from $[0,2\pi]$ to $[0, 2\pi N]$ due to the phase locking in Eq.\,(\ref{eq:lockingSUSY}).
Then, by remembering that the kinetic term of $\rho$ is mainly originated not from $S$'s but from $\phi$'s,
and hence, the canonically normalized inflaton multiplet $A$ is given by $A \sim \rho/v_\phi $,
%from which
the effective decay constant appearing in Eq.\,(\ref{eq:supereff}) is $f \sim N\,v_\phi$.
In this way, we can successfully supersymmetrize the phase locking mechanism which provides 
an effective decay constant larger than the VEVs in the linearly realized $U(1)$ symmetric models, i.e.,
$f = O(N \times v_{\phi, S})$.

Now, let us discuss behaviors of the scalar fields during inflation.
As in the generic model of the chaotic inflation in supergravity~\cite{Kawasaki:2000yn}, 
$X$ obtains a non-zero $F$-term, $F_X$, during inflation, which provides the inflaton potential.
The scalar components of $X$ as well as the s-inflaton are, on the other hand, stabilized at their origins 
due to couplings to $X X^\dag$ in the Kahler potential, assuming that those couplings provide them with
positive squared masses of the order of the Hubble scale squared during inflation.
After inflation, $F_X$ vanishes as the inflaton oscillation decays.
In this way, our model realizes the natural inflation model in the same way as the non-SUSY models.
It should be emphasized that we have not required any $R$-symmetry breaking to realize the explicite
$U(1)$ breaking effective potential in Eq.\,(\ref{eq:supereff}).
Thus, the energy density of the universe during inflation which is proportional 
to the size of the explicite $U(1)$ braking is not related to $R$-symmetry breaking, i.e.~the size of the gravitino mass.
Therefore, in our model, the gravitino can be small
%as small as possible
even for 
high scale inflation, and hence, it is compatible with low scale supersymmetry breaking.

\begin{center}
\begin{figure}[t]
 \begin{minipage}{.49\linewidth}
  \includegraphics[width=\linewidth]{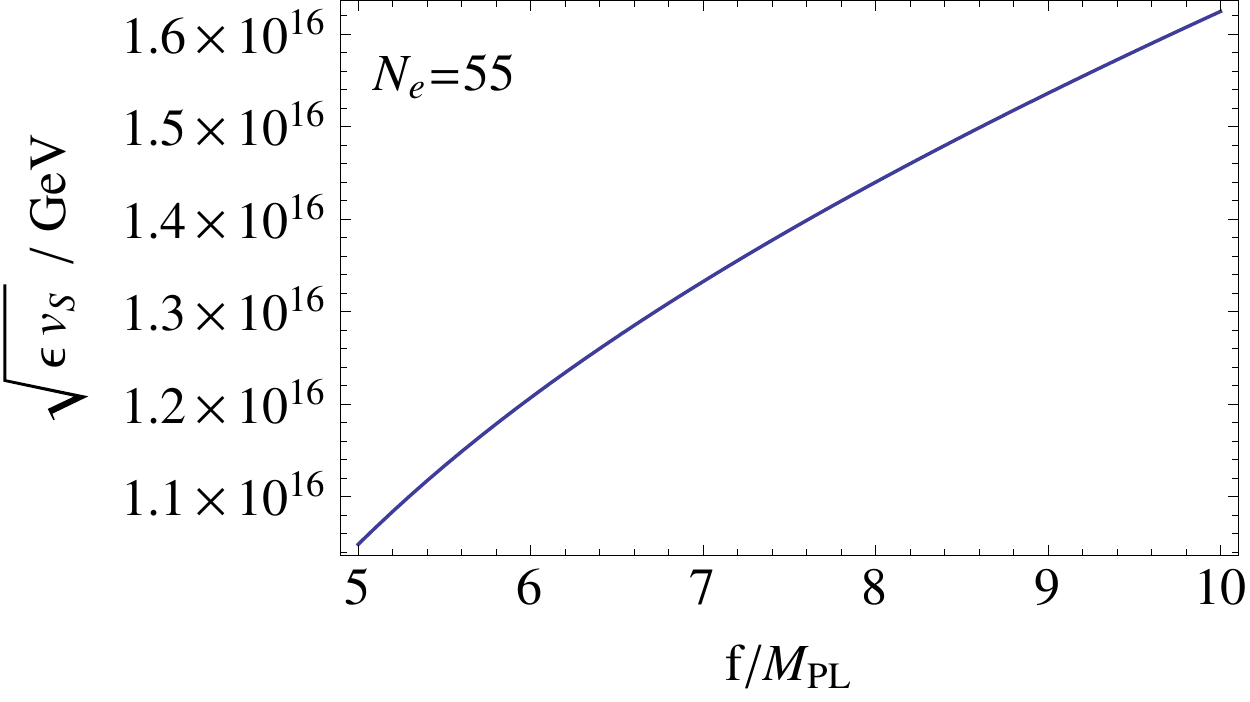}
 \end{minipage}
 \begin{minipage}{.49\linewidth}
  \includegraphics[width=1.0\linewidth]{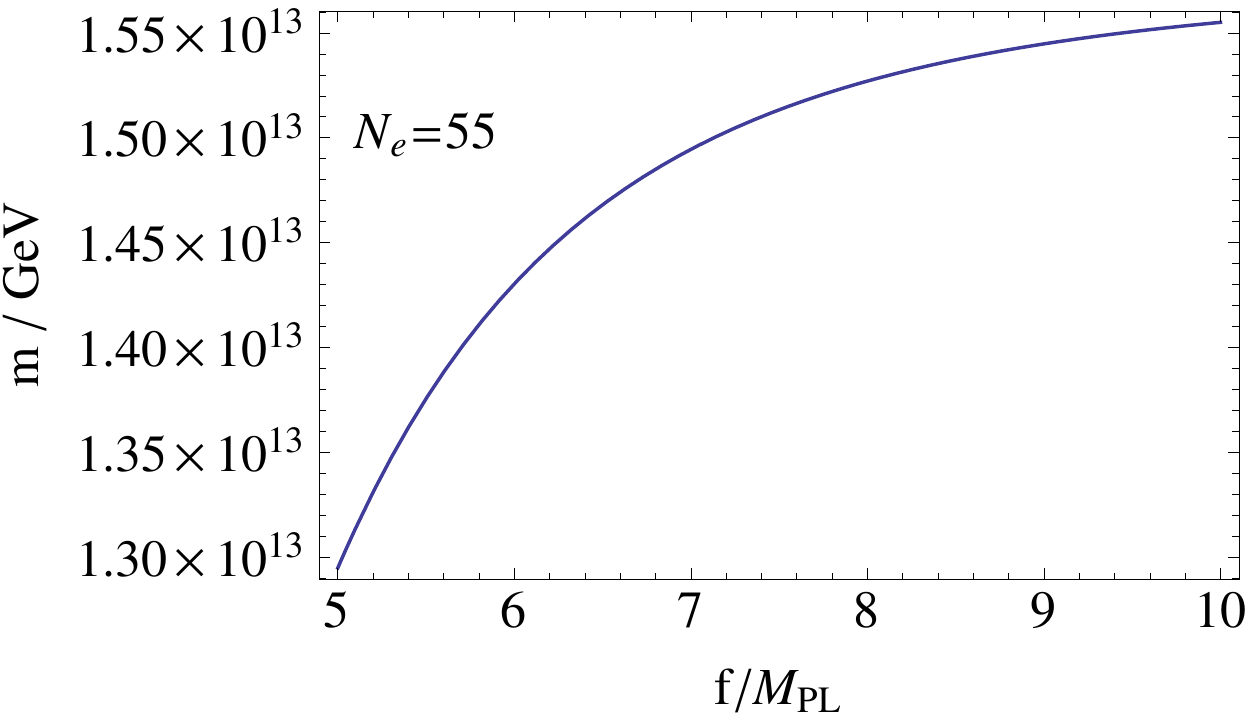}
 \end{minipage}
\caption{\sl \small
(Left) The energy scale $\sqrt{\epsilon v_S}$ for a given decay constant $f$.
(Right)The mass of the inflaton, $m$, for a given decay constant $f$.
}
\label{fig:mass}
\end{figure}
\end{center}

Finally, let us show the inflaton potential explicitly.
Since we assume that the VEVs of $\phi$'s and $S$'s are below the Planck scale,
the relevant Kahler potential terms of $\rho$ originate from,
\begin{eqnarray}
K &=& \phi \phi^\dag + \bar{\phi} \bar{\phi}^\dag +  S S^\dag + \bar{S} \bar{S}^\dag \nonumber \\
&=& v_\phi^2 (e^{\rho + \rho^\dag} +e^{-(\rho + \rho^\dag)}) + v_S^2 ( e^{(\rho + \rho^\dag)/N} + e^{-(\rho + \rho^\dag)/N}) \nonumber \\
&\simeq & v_\phi^2 (e^{\rho + \rho^\dag} +e^{-(\rho + \rho^\dag)})\nonumber \\
&=& v_\phi^2 + v_\phi^2 (\rho + \rho^\dag)^2 + \cdots
\end{eqnarray}
where we have assumed $v_\phi \sim v_S$ and $N\gg 1$.
Around the origin of the s-inflaton, the canonically normalized pseudo-NG multiplet $A$ is given by
\begin{eqnarray}
\label{eq:canonical}
A = \sqrt{2} v_\phi \rho \ ,
\end{eqnarray}
as expected.
Then, from Eqs.~(\ref{eq:supereff}) and (\ref{eq:canonical}), the potential of the inflaton is given by
\begin{eqnarray}
V(a) = 2 \epsilon^2 v_S^2 \left(
1 - \cos\left( \frac{2a}{N v_\phi} + \frac{\pi}{N} \right)
\right)\ ,
\end{eqnarray}
where $a = \sqrt{2}{\rm Im}A$.
The effective decay constant is
\begin{eqnarray}
f = N v_\phi/2,
\end{eqnarray}
which is far larger than the VEV of $\phi$ if $N\gg1$.
In Fig.~\ref{fig:mass}, we show the energy scale $\sqrt{\epsilon v_S}$ and the mass of the inflaton $m = \sqrt{2} \epsilon v_S / f $ for a given $f$~\cite{Freese:1990rb}.
Here, we have used the observed magnitude of the curvature perturbation, ${\cal P}_\zeta \simeq 2.2\times 10^{-9}$~\cite{Ade:2013uln},
and assumed that the number of e-foldings at the pivot scale of $0.002~{\rm Mpc}^{-1}$ is $55$.
%%%%%

\subsection{Reheating Processes}
In this subsection, let us discuss the reheating process in our set up.
For a while, we assume that the $Z_2\times Z_{2C}$ symmetry introduced in the previous section
is also respected in the interactions for the reheating process.
Here, it should be noted that the inflaton possesses a $Z_2$ odd parity around the vacuum, 
i.e. at $F_X = 0$;
\begin{eqnarray}
\vev{\rho} = i \pi \left(\frac{N (2k+1) -1}{2}\right),~~\vev{\sigma} = i \pi \left(k + \frac{1}{2}\right)~~(k = 0,1)\ .
\end{eqnarray}
That is, after $\phi$ and $S$ obtain non-vanishing VEVs, 
$Z_2\times Z_{2C}$ is broken down to a $Z_2$ symmetry.
This remaining $Z_2$ symmetry forbids the decay of the inflaton into 
gravitinos~\cite{Endo:2006zj,Nakamura:2006uc,Kawasaki:2006gs},
and hence, the model with this symmetry is free from the gravitino problem
induced by the inflaton decay.

Despite this advantage of the $Z_2$ symmetry, 
 the symmetry at the same time forbids the decay of the inflaton into the radiation composed of 
the Standard Model particles unless they are also charged under the $Z_2$ symmetry.%
\footnote{See also Ref.\,\cite{Harigaya:2014pqa} for related discussion.}
% or the $Z_2$ symmetry is explicitly broken.
Fortunately, however, it is not difficult to make the Standard Model fields have non-trivial charges under the symmetry.
For example, we may consider the charge assignment given in Table~\ref{tab:charge2},%
\footnote{
With the charge assignment given in Table 2, 
the bilinear term of the Higgs, $H_u H_d$ have the same charges as $S \bar{S}$ and $\phi \bar{\phi}$.
The super potential terms $W \supset Y_1 H_u H_d, Y_2 H_u H_d$ are allowed by the symmetry.
When the squared mass terms of the Higgs doublets ($m_{H_{u,d}}^2$), are smaller than those of $S$, $\bar{S}$,
$\phi$ and $\bar{\phi}$  ($m_{S,\phi}^2$), the vacuum with $| \vev{H_{u,d}}| \ll |\vev{S}|, |\vev{\bar{S}}| , |\vev{\phi}|, |\vev{\bar{\phi}}|$  is destabilized.
In order to avoid such a problem we assume either $m_{H_{u,d}}^2 > m_{S,\phi}^2$ or
somewhat suppressed couplings between $Y_{1,2}$ and $H_{u}H_{d}$.
The latter possibility can be realized, for example, by the so-called SUSY zero mechanism. 
}
in which we have promoted the $Z_2\times Z_{2C}$ symmetry to a $Z_4\times Z_{4C}$ symmetry
which has some advantageous features as we will see shortly.%
\footnote{Here, we have assigned vanishing $R$-charges to the Higgs doublets so that 
the model is consistent with the Pure Gravity Mediation model\,\cite{PMG} where the 
so-called $\mu$-term is generated from $R$-symmetry breaking~\cite{Inoue:1991rk,Casas:1992mk}.}
In the table, we have introduced right-handed neutrinos $N_R$ to explain the small mass of the neutrino 
by the seesaw mechanism~\cite{seesaw}.

\begin{table}[tdp]
\begin{center}
\begin{tabular}{|c|ccccccccc|}
\hline
& $\phi$ & $\bar{\phi}$ & $S$ & $\bar{S}$ & $Y_{1,2}$ & $Y_3$ & $X$ & $\epsilon$ & $\bar{\epsilon}$ \\
\hline
$U(1)$ & $+N$ & $-N$ & $+1$ & $-1$ & $0$ & $0$ & $0$ & $+1$ & $-1$ \\
$U(1)_R$ & $0$ & $0$ & $0$ & $0$ & $2$ & $2$ & $2$ & $0$ & $0$\\
$Z_4$ & $(-)^{N+1}$ & $(-)^{N+1}$ & $-$ & $-$ & $+$ & $-$ & $-$ & $+$ & $+$ \\
$Z_{4C}$ & $\phi\leftrightarrow \bar{\phi}$ &  & $S\leftrightarrow \bar{S}$ &   &&  & & $\epsilon \leftrightarrow \bar{\epsilon}$ &\\
\hline
& $\bar{u},Q,\bar{e}$ & $\bar{d},L$ & $H_u$ & $H_d$ & $N_R$ &  &  &  &  \\
\hline
$U(1)$ & $0 $ & $0$ & $0$ & $0$ & $0$ &  & &  &  \\
$U(1)_R$ & $1$ & $1$ & $0$ & $0$ & $1$ & & &  & \\
$Z_4$ & $i$ & $i$ & $-$ & $-$ & $i$ &  &  &  &  \\
$Z_{4C}$ & $i$ & $i$  & $-$ & $-$  & $ i $&  & & &\\
\hline
\end{tabular}
\end{center}
\caption{\sl \small Charge assignment of the inflaton sector and of the minimal supersymmetric standard model sector
$(\bar{u}, Q, \bar{e}, \bar{d}, L, H_u, H_d)$, and of the right-handed neutrinos $N_R$.}
\label{tab:charge2}
\end{table}

With the above charge assignment, the dominant decay mode is provided by the Kahler potential term,
\begin{eqnarray}
 K = \frac{y_1}{2 \mpl} X^\dag N_R N_R + {\rm h.c.},
\end{eqnarray}
which leads to the decay rate;
\begin{eqnarray}
\Gamma = \frac{y_1^2}{8\pi} \frac{m^3}{\mpl^2},
\end{eqnarray}
where $y_1$ is a constant.
The rehearing temperature is 
\begin{eqnarray}
T_{\rm RH} &\equiv& \sqrt{\frac{90}{\pi^2g_*}}\sqrt{\Gamma\mpl} \nonumber\\
&=& 1.6 \times 10^9~{\rm GeV} \times  y_1 \left(\frac{m}{1.5\times 10^{13}~{\rm GeV}}\right)^{3/2} \left(\frac{g_*}{200}\right)^{-1/2}\ ,
\end{eqnarray}
where $g_*$ is the effective degree of freedom of radiation.
For $y_1 = O(1)$, the reheating temperature is high enough for successful thermal leptogenesis~\cite{Fukugita:1986hr,Buchmuller:2005eh}.
Since the inflaton directly decays into the right-handed neutrino, non-thermal leptogenesis~\cite{Kumekawa:1994gx} 
is also possible.

As an interesting feature of the promoted discrete symmetry, $Z_4\times Z_{4C}$, 
the masses of the right-handed neutrinos are interrelated to the explicit $U(1)$ breaking parameter $\e$.
That is, due to $Z_4\times Z_{4C}$, the masses of the right-handed neutrinos 
are generated from the following coupling to the inflaton sector,
\begin{eqnarray}
W &=& \frac{y_2}{2\mpl} (\bar{\epsilon} S - \epsilon \bar{S} ) N_R N_R\ ,
% \nonumber \\
% & = & \frac{y}{2\mpl} \epsilon v_S N_RN_R \left(
%{\rm exp}\left[ \frac{b}{N} + i \frac{\pi}{2N} \right] -  {\rm exp}\left[-\frac{b}{N} - i \frac{\pi}{2N} \right] \right),
\end{eqnarray}
where $y_2$ is a constant. 
As a result, the right-handed neutrino masses are given by,
\begin{eqnarray}
M_R = 2 y_2 \frac{\epsilon v_S}{\mpl} = 1.4\times10^{14}~{\rm GeV}\times y_2 \frac{\epsilon v_S}{1.7\times 10^{32}~{\rm GeV}^2},
\end{eqnarray}
which are appropriate for the seesaw mechanism~\cite{seesaw}.

Let us also comment on the reheating process  when the model does not respect
the $Z_{2}\times Z_{2C}$ symmetry.
In this case, the inflaton  decays into other particles through the supergravity effect without introducing any 
particular couplings between  the inflaton and the Standard Model fields~\cite{Endo:2006qk,Endo:2007ih}.
On top of those spontaneous decays,  the inflaton may decay, for example, via 
the superpotential term,
\begin{eqnarray}
W =  y_3X H_u H_d
\end{eqnarray}
where $y_3$ is a constant, which tends to result in a very high reheating temperature.%
\footnote{If $y_3$ is larger than $m/\mpl$, inflation ends not by the fast roll of the inflaton but by the ``waterfall" 
of the higgs multiplets~\cite{Harigaya:2014roa}. In this case, the predictions on the spectral index and the tensor fraction are different from that of the standard natural inflation.}
It should be noted that without the $Z_{2}\times Z_{2C}$ symmetry, the inflaton in general decays into gravitinos, 
which leads to overproduction of the lightest supersymmetric particle and/or spoils the success of the 
Big-Bang Nucleosynthesis. 
To evade this inflaton induced gravitino problem, we need to assume either the discrete 
symmetries as we did in this paper or to assume the fields in  the SUSY breaking sector 
much heavier than  the inflaton~\cite{Nakayama:2012hy}.

\section{Discussion and Conclusions}
In this paper, we gave detailed descriptions of ``phase locked" inflation
where the effectively trans-Planckian decay constant for natural inflation
is achieved by the ``phase locking" mechanism\,\cite{Harigaya:2014eta}.
As we have discussed,  the effectively trans-Planckian decay constant 
originates from the charge assignments of fields which break the $U(1)$ symmetry. 
This mechanism should be contrasted with other attempts to realize the effectively trans-Planckian 
decay constant field variation by,  for example, alignment between several potentials of natural inflation
or by using the collective behaviour of multi-inflatons of natural inflation. 
In our model, on the other hand, the effectively trans-Planckian decay constant 
can be achieved by only two fields without having alignments.

We also construct a model of supersymmetric natural inflation based 
on the ``phase locking" mechanism.
The advantageous feature of our model is that the explicit symmetry breaking of the $U(1)$ symmetry
is separated from the $R$-symmetry breaking, and hence, the model does not 
lead to a large gravitino mass even for a high inflation energy scale.
Therefore, our model is compatible with low scale supersymmetry. 

We have also discussed how the reheating process proceeds in this model.
In particular, we found that the decay of the inflaton can be well controlled by discrete symmetries
which forbids the decay modes into gravitnos.
We also found that the masses of the right-handed neutrinos appropriate for the seesaw mechanism
can be related to the inflation scale  in a certain class of  models with discrete symmetries.

\section*{Acknowledgements}
We thank Kawasaki Masahiro and Tsutomu T. Yanagida for useful discussion.
This work is supported by
Grant-in-Aid for Scientific research 
from the Ministry of Education, Science, Sports, and Culture (MEXT), Japan, No. 24740151 
and 25105011 (M.I.), from the Japan Society for the Promotion of Science (JSPS), No. 26287039 (M.I.),
the World Premier International Research Center Initiative (WPI Initiative), MEXT, Japan (K.H. and M.I.)
and a JSPS Research Fellowships for Young Scientists (K.H.).

\appendix

\section{Tunneling between valleys}

\begin{figure}[tb]
\begin{center}
\begin{minipage}{.49\linewidth}
  \includegraphics[width=\linewidth]{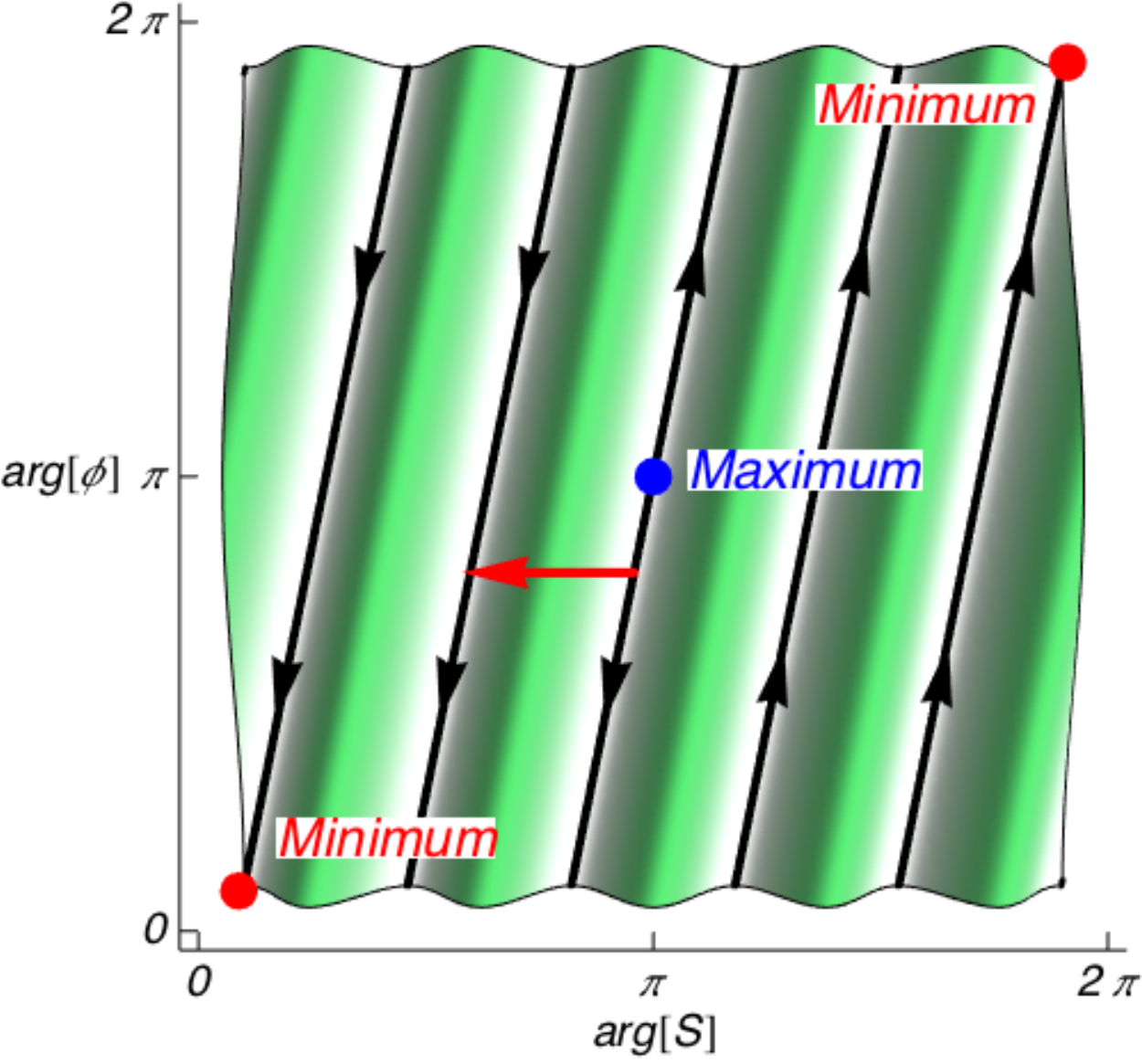}
 \end{minipage}
 \begin{minipage}{.49\linewidth}
  \includegraphics[width=1.0\linewidth]{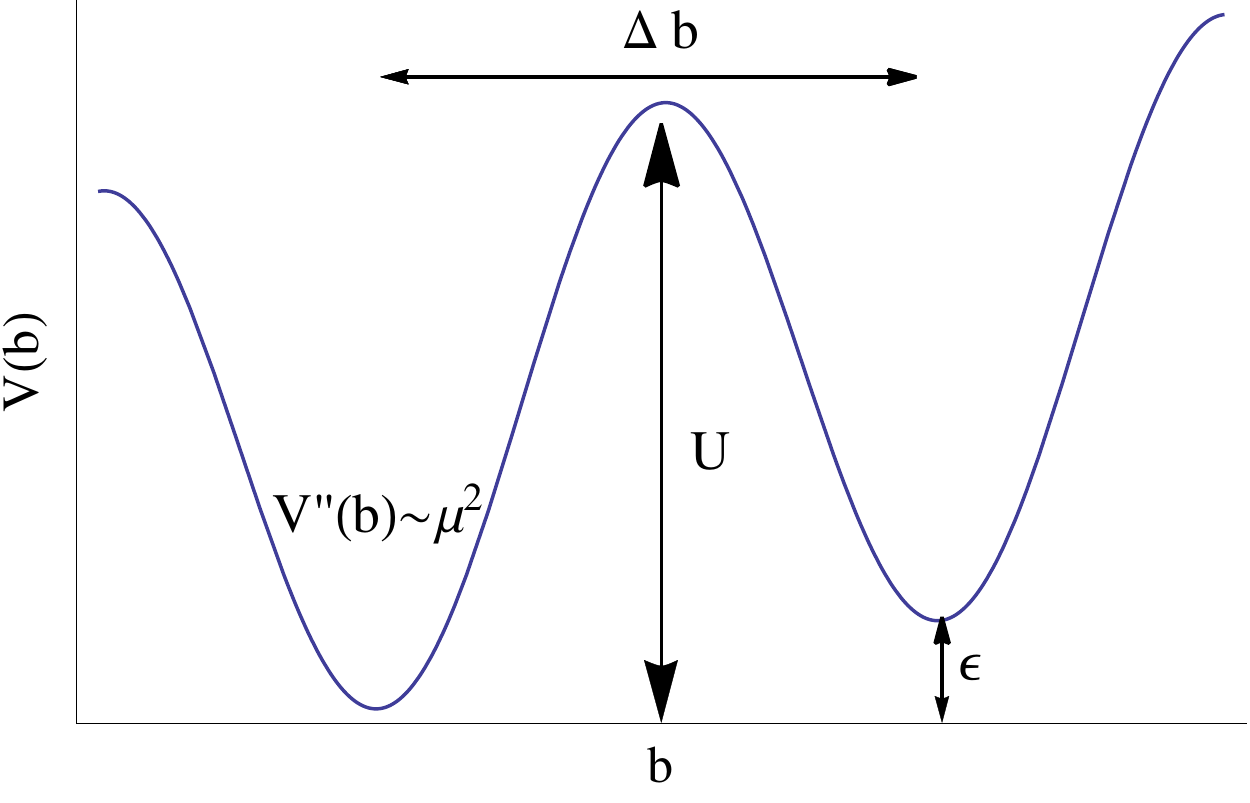}
 \end{minipage}
\caption{\sl \small
(Left) Same as Fig.~\ref{fig:lock}, with a possible tunneling process indicated by the red arrow.
(Right) The potential along the tunnelling trajectory.
Here, $b$ denotes the canonically normalized field along the trajectory.
}
\label{fig:tunneling}
\end{center}
\end{figure}

In this section, we estimate the tunneling rate between the valleys of the inflaton potential, whose trajectory is shown in Fig.~\ref{fig:tunneling} by the red arrow.
For a large $N$, distance between valleys is short, 
and hence the tunneling process might disturb the inflation dynamics.
To be concrete, we consider the two field model presented in section~\ref{sec:locking}.
The potential along the tunneling trajectory is sketched in the right panel of Fig.~\ref{fig:tunneling},
with the parameters
\begin{eqnarray}
\Delta b &=& \frac{2\pi}{N^2} f \sim \frac{2\pi}{N} |\vev{\phi}|, \nonumber  \\
\epsilon &\sim&  \frac{2\pi}{N}\Lambda^4, ~~ \Lambda^4 \sim 10^{-8}M_{\rm pl}^4  \nonumber\\
U &\sim& \frac{1}{M_{\rm Pl}^{N-3}} |\vev{\phi}| |\vev{S}|^N, \nonumber \\
\mu^2 &\sim& \frac{1}{M_{\rm Pl}^{N-3}} |\vev{\phi}| |\vev{S}|^N \frac{N^2}{|\vev{\phi}|^2}.
\end{eqnarray}
Here, $b\sim {\rm arg}S /N \times f $ denotes the canonically normalized field along the trajectory.

Note that $U > \epsilon$ is required so that $b$ does not move classically, which gives lower bound on $\vev{\phi}$ and $\vev{S}$.
Under this assumption, the size of the core of the bounce solution of the tunneling process in the thin wall approximation~\cite{Coleman:1977py},
$U^{1/2} \Delta b / \epsilon$, is larger than the thickness of the skin of the bounce solution, $\mu^{-1}$.
Thus the thin wall approximation is applicable.
The tunneling rate per unit volume per unit time is given by~\cite{Coleman:1977py}
\begin{eqnarray}
\Gamma  &\sim& \mu^4 {\rm exp} (- S_0),\nonumber \\
S_0 &\sim& \frac{27}{2 \epsilon^3} \pi^2 U^2 \Delta b^4 = \frac{27\pi^6}{N} \frac{|\vev{\phi}|^6 |\vev{S}|^{2N}}{\Lambda^{12}M_{\rm Pl}^{2N-6}}.
\end{eqnarray}
Since $\Lambda \sim 10^{-2} M_{\rm Pl} \ll M_{\rm Pl} $, $S_0$ is far larger than one for sufficiently large but sub-Planckian $\vev{\phi}$ and $\vev{S}$.
The tunneling process shown in Fig.~\ref{fig:tunneling} can be suppressed.
We note that the lower bound on $\vev{\phi}$ and $\vev{S}$ from the conditions $U/\epsilon$ and $S_0 \gg 1$ is relaxed by the generalization discussed at the end of section~\ref{sec:locking}.

\end{document}